%% file: main.tex
\documentclass[a4paper,fleqn]{cas-dc}
\ExplSyntaxOn
\RenewDocumentCommand \printorcid {}{
  \seq_if_empty:NF \g_stm_orcid_seq {
    \group_begin:
      \tex_let:D \thefootnote \relax \footnotetext
      {
        \raggedright
        \textsc{orcid}(s):\c_space_token
        \seq_use:Nn \g_stm_orcid_seq { ;~ }
      }
    \group_end:
  }
}
\ExplSyntaxOff
\usepackage[authoryear,round]{natbib}
\makeatletter
\renewcommand{\ttdefault}{lmtt}
\makeatother
\usepackage[final]{microtype} 
\usepackage{graphicx}
\usepackage{tikz}
\usetikzlibrary{automata,positioning,arrows.meta}
\usepackage{enumitem}
\usepackage{subcaption}
\usepackage{xurl}
\usepackage{ragged2e}
\usepackage{algorithm}
\usepackage{algpseudocode}
\usepackage{xcolor}

\algrenewcommand\algorithmicrequire{\textbf{Inputs:}}
\algrenewcommand\algorithmicensure{\textbf{Output:}}
\usepackage{listings}
\lstdefinelanguage{SPARQL}{
morekeywords={PREFIX,SELECT,CONSTRUCT,ASK,DESCRIBE,WHERE,FILTER,OPTIONAL,GRAPH,UNION,SERVICE,FROM,NAMED,ORDER,BY,LIMIT,OFFSET,BIND,VALUES},
  sensitive=true,
  morecomment=[l]{\#},
  morestring=[b]"
}
\AtBeginEnvironment{itemize}{\RaggedRight} % weniger Zwang zum Blocksatz in Listen

\def\tsc#1{\csdef{#1}{\textsc{\lowercase{#1}}\xspace}}
\tsc{WGM}
\tsc{QE}
\tsc{EP}
\tsc{PMS}
\tsc{BEC}
\tsc{DE}
\begin{document}
\let\WriteBookmarks\relax
\def\floatpagepagefraction{1}
\def\textpagefraction{.001}

% Short title
\shorttitle{Using LLM Agents for Fault-Tolerant Control}

%Old titleMethodological Framework for Zeroshot Fault Handling in Process Plants with LLM Agents

% Short author
\shortauthors{J Vyas et~al.}

% Main title of the paper
\title [mode = title]{From Detection to Action: Using LLM Agents for Fault-Tolerant Control}                      
% Title footnote mark
% eg: \tnotemark[1]
% \tnotemark[1,2]

% Title footnote 1.
% eg: \tnotetext[1]{Title footnote text}
% \tnotetext[<tnote number>]{<tnote text>} 
\nonumnote{\noindent\textbf{Acknowledgment:} This research [project ProMoDi] is funded by dtec.bw – Digitalization and Technology Research Center of the Bundeswehr. dtec.bw is funded by the European Union – NextGenerationEU.\\
Financial support from ABB for the Autonomous Industrial Systems Laboratory at Imperial College London is gratefully acknowledged.\\
The first two authors contributed equally to this work.}

% First author
%
% Options: Use if required
% eg: \author[1,3]{Author Name}[type=editor,
%       style=chinese,
%       auid=000,
%       bioid=1,
%       prefix=Sir,
%       orcid=0000-0000-0000-0000,
%       facebook=<facebook id>,
%       twitter=<twitter id>,
%       linkedin=<linkedin id>,
%       gplus=<gplus id>]

\author[1]{Javal Vyas}
\credit{Writing - original draft, Conceptualization, Methodology, Investigation, Software, Validation}
\cormark[1]
\ead{j.vyas24@imperial.ac.uk}

\author[2]{Milapji Singh Gill}
\credit{Writing - original draft, Conceptualization, Methodology, Investigation, Software, Validation}
\author[1]{Artan Markaj}
\credit{Writing - review \& editing, Software}
\author[2]{Felix Gehlhoff}
\credit{Review-Editing, Funding, Supervision}
\author[1]{Mehmet Mercangöz}
\credit{Review-Editing, Funding, Supervision}

%  Credit authorship

% Address/affiliation
\affiliation[1]{organization={Imperial College London}, 
    city={London},
    % citysep={}, % Uncomment if no comma needed between city and postcode
    % postcode={1043 NX}, 
    % state={},
    country={United Kingdom}}

%\credit{Data curation, Writing - Original draft preparation}

% Address/affiliation
\affiliation[2]{organization={Helmut Schmidt University},
    % addressline={}, 
    city={Hamburg},
    % citysep={}, % Uncomment if no comma needed between city and postcode
    postcode={22043}, 
    country={Germany}}

% Corresponding author text
%\cortext[cor1]{Corresponding author}
\cortext[cor2]{Corresponding author}

% Footnote text
% \fntext[fn1]{This is the first author footnote. but is common to third
%   author as well.}

% For a title note without a number/mark
% \nonumnote{This note has no numbers. In this work we demonstrate $a_b$
%   the formation Y\_1 of a new type of polariton on the interface
%   between a cuprous oxide slab and a polystyrene micro-sphere placed
%   on the slab.
%   }

% Here goes the abstract
\begin{abstract}
We propose an agentic Large Language Model (LLM) framework for active Fault-Tolerant Control (FTC) that transforms fault detection outputs into constraint-aware recovery actions grounded in plant-specific knowledge. The approach couples (i) a multi-agent workflow that decomposes operator duties into monitoring, planning, action synthesis, simulation, validation, and reprompting; (ii) a Digital Process Plant Twin (DPPT) that exposes plant data, models, and a simulation service for pre-execution testing; and (iii) a Graph Retrieval-Augmented Generation (Graph RAG) layer built on the CPSMod ontology, which organizes plant knowledge (structure, function, hybrid dynamics, control context, and fault semantics) into a graph that supports relation-aware, multi-hop retrieval for the agents. Corrective actions are generated as minimal-risk state-machine recovery paths and corresponding discrete commands or continuous setpoint adaptations, then validated deterministically against interlocks, envelopes, and dynamic feasibility before any actuation. If no acceptable plan is found within a bounded time window, control is handed to a safety fallback. The framework is evaluated in simulation on two representative benchmarks: a discrete batch Mixing Module and a Continuous Stirred-Tank Reactor (CSTR) under closed-loop PID regulation. Results with lightweight LLMs (GPT-4o-mini and GPT-4.1-mini) show that semantically grounded agents can derive valid recovery decisions within latency budgets compatible with the respective process dynamics, demonstrating a practical pathway from detection to validated corrective action across both discrete and continuous FTC tasks.
\end{abstract}

% Use if graphical abstract is present
% \begin{graphicalabstract}
% \includegraphics{figs/grabs.pdf}
% \end{graphicalabstract}

% Research highlights
\begin{highlights}
\item A multi-agent LLM framework for active fault-tolerant control is proposed.
\item Semantic knowledge grounding enables constraint-aware corrective actions.
\item Simulation-based validation ensures safety before control execution.
\item The framework handles both discrete and continuous recovery tasks.
\item Effectiveness is demonstrated on mixing and CSTR case studies.

\end{highlights}

% Keywords
% Each keyword is seperated by \sep

\maketitle

\input{01_Introduction}

\input{03_StateOfTheArt}

\input{04_Framework}

\input{05_CaseStudies}

\input{06_Results}

\input{07_Discussion}

\input{08_SummaryAndOutlook}

\appendix

\printcredits

\section*{Declaration of Generative AI and AI-assisted technologies in the writing process}
During the preparation of this work the authors used Generative AI for minor text refinements.
After using this tool, the authors reviewed and edited the content as needed and take full responsibility
for the content of the publication.

\section*{Data availability}
The code necessary to generate the presented results is openly available under \url{https://github.com/AISL-at-Imperial-College-London/ctrl-alt-recover}. The reused standards-based ODPs and tools for their use are publicly maintained at \url{https://github.com/hsu-aut}.

\section*{Declaration of competing interest}
The authors declare that they have no known competing financial interests or personal relationships
that could have appeared to influence the work reported in this paper.

%% Loading bibliography style file
%\bibliographystyle{model1-num-names}
\bibliographystyle{cas-model2-names}

% Loading bibliography database
\bibliography{cas-refs}

\end{document}

%% file: 01_Introduction.tex
\section{Introduction} \label{Intro}

Despite advances in automation, many process industry operations remain heavily dependent on the expertise of human operators for accurate situation assessment and decision making under unforeseen conditions. This is particularly true where data is limited or system behavior deviates from expected patterns (\cite{BALDEA2025109064}). Process plants exhibit tightly coupled flows of energy, materials, and information, so even small disturbances can propagate across subsystems. Faults in process systems often manifest through subtle, state-dependent symptoms that hinder early detection and diagnosis (\cite{Webert.2022}), while alarm floods, where many alarms trigger in rapid succession, can overwhelm operators (\cite{Manca.2021}) and complicate timely decision making. In such situations, operators must interpret noisy and incomplete evidence, map it to plant topology and operating modes, and execute safe recovery procedures. This is compounded by uneven expertise across shifts and limited decision support (\cite{Markaj.2024}). These challenges motivate greater autonomy and drive new approaches in active Fault-Tolerant Control (FTC) for process plants. FTC aims to preserve safe, autonomous operation in the presence of faults by compensating their effects online (\cite{Olivier_etal_2017}). 

Over the years, various Artificial Intelligence (AI)-driven techniques have been proposed for active FTC. Especially Reinforcement Learning (RL) and agent-based methods have been explored as complements to classical process control. RL agents can autonomously optimize complex process tasks such as reactor operation and crystallization (\cite{Alhazmi_etal_2022, Manee2022, MDPI2024}). However, their broader adoption in process industries remains limited due to several fundamental challenges. In particular, RL approaches often exhibit low sample efficiency (\cite{MDPI2024, ACS2024}), limited generalizability across different systems (\cite{MDPI2024, CrystalGPT2023}), and require carefully designed reward functions to ensure stable and interpretable policy learning (\cite{MDPI2024, Manee2022}). Beyond RL, agent-based concepts are increasingly considered to replace or augment human operators (\citep{Lee_etal_2021}). While these approaches highlight the potential of AI-driven autonomy in process plants, extensive training, simulation and validation remain essential before deployment. 

Recent advances in Large Language Models (LLMs), together with agentic tooling, enable end-to-end orchestration of control actions without large task-specific datasets (\cite{Xia.11205539}). In the context of active FTC, LLM agents can coordinate condition monitoring, reason over plant-specific context for plant recovery, and invoke tools for pre-execution validation, thereby improving sample efficiency and operational safety (\cite{Gill_Vyas}). However, effective deployment requires grounding agents in plant-specific knowledge that is distributed across heterogeneous engineering artifacts (\cite{RUPPRECHT2026101209}) and must be integrated into a unified representation. Artifacts include P\&IDs, cause–effect matrices, interlock lists, operating envelopes, maintenance records and physics-based simulation models, which together encode structure, function, behavior and fault semantics (\cite{Gowaikar.17.12.2024,Gill_Vyas}). 
Moreover, process plants combine continuous and discrete behaviors that make timely reasoning difficult. While document-level Retrieval-Augmented Generation (RAG) helps with grounding, it is less suited when recovery requires explicit traversal of distributed relations among equipment, faults, interlocks, operating modes, and control actions (\cite{DanielOvalle.2025})
.

Thus, the following research questions (RQs) need to be addressed in the context of FTC:
\begin{itemize}
  \item \textbf{RQ1:} \textit{How should plant-specific knowledge be linked, represented and retrieved so that LLM agents are properly grounded for active FTC?}
  \item \textbf{RQ2:} \textit{How can LLM agents synthesize safe recovery plans with corrective control actions and validate them prior to execution?}
  \item \textbf{RQ3:} \textit{Can LLM agents derive effective corrective actions within a reasonable time span in both open-loop and closed-loop control scenarios?}
\end{itemize}

The following contribution is structured as follows. Sec. 2 reviews the state of the art on LLM-based engineering and control. Sec. 3 presents our LLM-based agent framework for active FTC and details its components. Sec. 4 evaluates the framework in two case studies: (i) a Modular Mixing Unit with discrete supervisory sequences and (ii) a Continuous Stirred-Tank Reactor with time-continuous dynamics. Sec. 5 reports results against predefined evaluation criteria. Sec. 6 discusses implications and limitations and answers the RQs. Sec. 7 concludes with a summary and an outlook.

%% file: 03_StateOfTheArt.tex
\section{State of the art} \label{StoA}

\subsection{LLMs for process operations} \label{Eng}
LLMs have moved into the scope for engineering applications as advances in natural-language processing, code generation, and multimodal reasoning mature. A recent overview by \cite{RUPPRECHT2026101209} highlights how LLMs are increasingly used to bridge unstructured engineering knowledge (e.g., manuals, reports, logs) and structured engineering workflows, supporting tasks in engineering design, system analysis, and operational decision support.

Across these application areas, end-to-end workflows typically rely on three complementary technological paradigms. First, RAG grounds LLM outputs in external information sources to mitigate outdated training knowledge and hallucinations. Graph-based variants such as Graph RAG further enable relation-aware, multi-hop retrieval across dispersed facts (\cite{srinivas2025autochemschematicaiagenticphysicsaware}). Second, agentic architectures embed LLMs as reasoning components within orchestrated workflows, with specialized agents for tasks such as monitoring, validation, and reporting, to integrate heterogeneous tools and data sources (\cite{Vyas.2025}). Third, planning capabilities support the generation and evaluation of action sequences under constraints, enabling structured, goal-directed problem solving (\cite{jobs2025benchmarkplanningcontrollarge}).

In the context of engineering design, LLM agents have been applied to assist with the analysis and modification of process flow designs. For example, \cite{Lee.2024} developed a multi-agent system that takes an existing process flow diagram (PFD) as input and generates suggestions for process improvements or alternative designs. By parsing the flowsheet structure and accessing external literature via tools, the agents proposed modifications such as heat integration steps or alternative separation sequences.
Similarly, \cite{Gowaikar.17.12.2024} demonstrated how LLM-based agents can generate engineering diagrams from natural-language descriptions. Their multi-step copilot translates textual process descriptions into structured P\&IDs, which can subsequently be reviewed and corrected by human experts, enabling iterative refinement.

Beyond design, LLMs have been leveraged for analysis and knowledge extraction from heterogeneous engineering data. \cite{DanielOvalle.2025} constructed knowledge graphs from academic literature in the process systems engineering domain and employed an LLM agent with Graph RAG to synthesize information for answering optimization-related queries. Such approaches illustrate the potential of LLMs to formalize dispersed knowledge and support analytical reasoning over complex process information.

LLM agents have also been explored as operator assistants and decision-support systems. \cite{Sakhinana.2024} proposed a centralized LLM-based agent capable of answering ad-hoc operator queries and generating fragments of automation or control code on demand, using a Graph RAG mechanism to retrieve relevant documentation and code examples from a company knowledge base. At a higher operational level, \cite{Pajak.2025} employed a multi-agent system to balance economic and environmental objectives in a gas–oil separation plant, where LLM agents interacted with simulation tools (e.g., Aspen HYSYS) to evaluate and negotiate operational trade-offs. 

\subsection{LLMs for control tasks} \label{Control}
While the focus of this work is on process systems, it is instructive to consider how LLM-based control concepts are explored in other Cyber-Physical System (CPS) domains. Across areas such as building automation, robotics, and manufacturing, LLMs are investigated for control-related tasks at different levels of abstraction, ranging from direct action selection to high-level planning and supervisory coordination. Many of the underlying challenges, e.g. uncertainty, changing operating conditions, and the need for safe decision making, are shared across these domains.
 
In some studies, LLMs are used directly for control action selection. For example, the authors of this work have previously introduced an LLM-driven multi-agent framework for an industrial temperature regulation task, where agents monitored sensor readings and adjusted actuator settings in real time (\cite{Vyas.2025}). Control decisions were generated through natural-language reasoning and validated against expected system behavior, illustrating how LLMs can be embedded directly into control loops.
Similarly, in the context of building automation, \cite{HVACLLM} demonstrated that an LLM can function as a controller for HVAC systems. At each time step, the LLM received a structured prompt describing the control objective, selected state–action examples, and current sensor readings, and responded with a control action such as adjusting a thermostat or fan setting. Despite minimal task-specific tuning, the LLM-based controller achieved performance comparable to a RL controller and showed robustness across varying weather and occupancy conditions.

% In robotics, LLMs are more commonly employed as high-level planners rather than low-level controllers. A prominent example is Google’s PaLM-SayCan framework (\cite{ahn2022icanisay}), which combines an LLM with a set of predefined robotic skills and an affordance function that filters feasible actions based on the current environment. Given a user goal, the LLM decomposes the task into sequential steps in natural language, while the affordance model ensures that only physically executable actions are selected. This separation of high-level reasoning and low-level execution enables flexible task planning while maintaining safety and feasibility. Extensions such as PaLM-E further integrate visual sensor data directly into multimodal LLMs (\cite{Driess.362023}), allowing the model to reason jointly over perception and prior knowledge when guiding robotic behavior.

In manufacturing and industrial automation, LLMs are increasingly explored as supervisory or orchestration layers. For instance, \cite{Xia.11205539} proposed an architecture in which an LLM operates above a hierarchical automation system, receiving textualized event streams from different levels of a production line and generating coordinated control or scheduling actions. In this role, the LLM does not replace existing controllers but supervises and adapts their interaction, with digital twins and simulation models providing situational awareness and evaluation of proposed actions.

\subsection{Contributions of this study}
Recent work shows that LLMs can support process operation workflows by querying documentation, calling tools, and generating automation or control code, effectively bridging natural-language intent and machine execution. However, these capabilities do not yet amount to deployable fault handling in safety-critical process plants. Three gaps recur across the literature: 

\begin{enumerate} [nosep,leftmargin=*]
    \item (i) \emph{Grounding}: plant knowledge is fragmented across heterogeneous artifacts, and text-only retrieval often misses the \emph{relationships} needed to derive corrective actions.
    \item (ii) \emph{Assurance}: actions are rarely accompanied by auditable traces and explicit feasibility checks against interlocks, operating envelopes, and process physics before execution.
    \item (iii) \emph{Timeliness/Robustness}: recovery must operate under hybrid dynamics, partial observability, and distribution shift, where plausible but delayed or incorrect actions can be unsafe.
\end{enumerate}

We address these gaps with three contributions. First, we enable relation-aware, multi-hop grounding via an ontology and Graph RAG to retrieve actionable dependencies linking equipment, control logic, operating modes, and fault semantics. Second, we enforce safety through minimal, explainable recovery plans with deterministic pre-execution validation (interlocks, envelopes, and dynamic feasibility) and an auditable decision trace, implemented via a multi-agent workflow (planning, action synthesis, simulation, validation, reprompting). Third, we support bounded-latency operation through tool-assisted monitoring, constrained action generation, runtime verification, and a safety fallback if no admissible plan is found. The resulting framework (see Sec.~\ref{Framework}) connects fault detection outputs to validated corrective actions for hybrid process systems.

%% file: 04_Framework.tex
\section{LLM-based agentic framework for fault-tolerant control} \label{Framework}

\subsection{Overview}
\begin{figure*}[pos=t]
    \centering
    \includegraphics[width=1\textwidth]{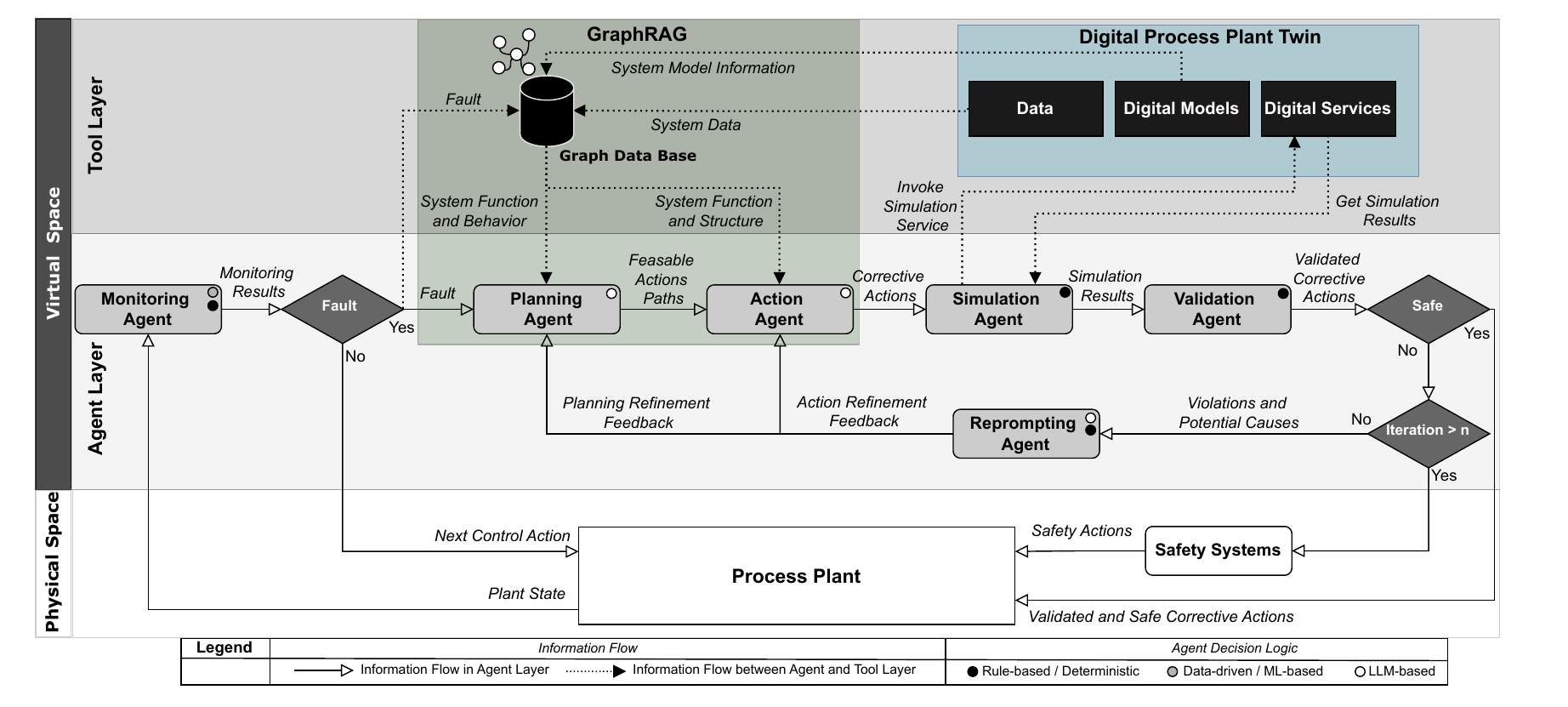}
    \caption{\textbf{Framework for active FTC, structured into Physical Space and Virtual Space with Agent Layer and Tool Layer.}}
    \label{fig:framework}
\end{figure*}

The following Sec. \ref{Framework} provides an overview of the framework used to integrate LLM agents into active FTC. We distinguish two spaces, the Physical and the Virtual one (see Figure \ref{fig:framework}). The Physical Space contains the Process Plant and a Safety System that executes a fail-safe strategy if no valid corrective action is found within a bounded time window. The Virtual Space is split into a \textit{Tool Layer} (see Sec. \ref{Tool}) and an \textit{Agent Layer} (see Sec. \ref{Agent}). Two types of information flows are distinguished: firstly, flows between the Physical Space and the \textit{Agent Layer}, as well as among the agents themselves, which are depicted with solid lines. Secondly, flows between the \textit{Agent Layer} and the \textit{Tool Layer}, which are depicted with dashed lines. Moreover, the agent decision logic is categorized into \textit{rule-based/deterministic}, \textit{data-driven/ML-based}, and \textit{LLM-based}, as indicated in the legend.

\subsection{Tool layer} \label{Tool}
The \textit{Tool Layer} exposes two shared components. Firstly, a Digital Process Plant Twin (DPPT), which is an executable tool stack that provides plant data, physics-/data-driven models, and validation services to automate FTC tasks. Secondly, an ontology that integrates heterogeneous engineering artifacts into a machine-interpretable representation to ground LLM agents via structured, multi-hop prompt augmentation and context retrieval across structure, function, behavior, condition monitoring as well as fault information.

\paragraph{\textbf{Digital Process Plant Twin:} \label{DPPT}}
We ground our notion of a DPPT in two complementary perspectives from the literature. Following the five-part view by \cite{Tao.2019}, a digital twin comprises the physical entity, its virtual counterpart, their bidirectional connection, a data space, and digital service capabilities. In parallel, we adopt the definition introduced by \cite{Kritzinger.2018}, which includes a digital model (no automatic coupling), a digital shadow (automatic one-way synchronization from physical to digital), and a digital twin (automatic two-way synchronization). The digital twin capabilities should be scoped to the decision problem and calibrated in fidelity, coverage, and services to meet the accuracy and latency requirements of the task (\cite{Reinpold.2024}, \cite{Gill.2022}). Because we target active FTC, our DPPT instantiates only the data, models, and services needed to detect anomalies, localize faults, synthesize constraint compliant corrective actions, and validate them within prescribed latency budgets. In this configuration, the DPPT (blue rectangle in Figure \ref{fig:framework}) serves as the callable tool layer for LLM-driven agents and supports condition monitoring and simulation based validation.

Operational data in the DPPT comprise time-synchronized sensor and actuator streams, including level, flow, pressure, temperature, valve states, and pump speed. They further include batch and sequence logs as well as alarms and events. Moreover, outputs from condition monitoring, such as detected anomalies and faults  that are generated by data driven models, are stored. Static engineering data include P\&IDs, parameter limits, operating envelopes, cause-and-effect matrices and the hierarchical plant breakdown structure.

The DPPT maintains two types of digital models. The first type is the simulation model of the plant, which encodes structure, function, and dynamics at the granularity required to evaluate generated corrective control actions. Structural and functional descriptions capture equipment connectivity and unit roles. Behavioral abstractions capture supervisory modes and sequences as finite state machines. Physics-based and reduced order models implement the continuous time mass and energy balances, constraints and actuation effects.The second type comprises data driven models for condition monitoring and fault detection. 

\paragraph{\textbf{Graph RAG:}} \label{GraphRAG}
In order to address RQ1, we use an ontology to integrate heterogeneous data and to represent the system in a machine interpretable way, enabling LLM agents to be grounded in plant specific knowledge. We distinguish between the terminological level of ontologies, the T-Box, which defines the schema, and the assertional level, the A-Box, which contains the instance data for a specific plant. This separation lets us reuse the same schema for different assets while populating it with plant-specific instances for active FTC. Additionally, it enables semantic harmonization and contextualization: the T-Box assigns explicit meaning to the data by providing a shared vocabulary and formal relations, allowing the integrated information to be both machine- and human-interpretable.

In order to depict the terminological knowledge of the knowledge graph, the modular alignment ontology \texttt{CPSMod} is reused. The ontology is modular, and its core concepts and relations are derived from industrial standards in order to maximize reuse and ensure a consistent vocabulary across projects. The method introduced by \cite{Hildebrandt.2020} was applied to model the ontology with different reusable standards-based ontology design patterns (ODPs). The ontology is specified using the W3C stack, namely RDF and OWL for modeling and is additionally annotated with natural-language metadata (e.g., \texttt{rdfs:label} and \texttt{rdfs:comment}) to make classes, relations and instances interpretable in text form. Prior studies (e.g., \cite{Reif}) indicate that such annotations can improve LLM-based interpretation of knowledge graph content.

\begin{figure*}[pos=t]
    \centering
    \includegraphics[width=1\textwidth]{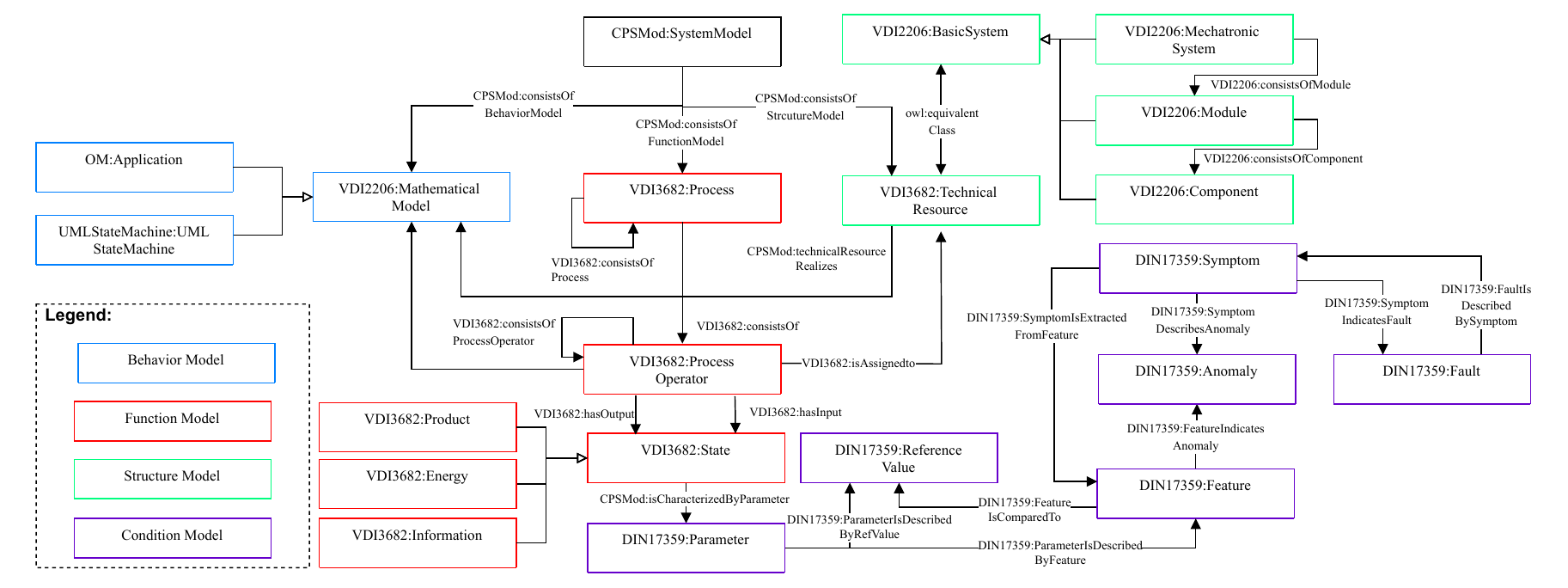}
    \caption{\textbf{Excerpt from the CPSMod alignment ontology integrating five standards-based ODPs.}}
    \label{fig:ontology}
\end{figure*}

An excerpt of the model with information relevant for the task is depicted as an unified modeling language (UML) class diagram in Figure \ref{fig:ontology}. This ontology, already introduced and applied by the authors in previous works (\cite{Gill11205677, Gill.2024}) for other CPS from the process and the aviation domain, provides unified semantics with regard to the domain knowledge required. It connects structure (green classes), function (red classes), behavior (blue classes), as well as condition monitoring and fault (purple classes) information and therefore supports relation aware, multi hop retrieval. \texttt{CPSMod} links those classes from different ODPs by using object properties (e.g., \nolinkurl{owl:equivalentClass}, \nolinkurl{rdfs:subClassOf}, new object properties)  so that agents can traverse multi-hop reasoning paths of specific subgraphs. ODPs included are:

\begin{itemize} [nosep,leftmargin=*]
    % \item ODP \textit{DIN 77005} for a unified semantic regarding lifecycle data of the plant \nolinkurl{(DIN77005:LifeCycleRecord, DIN77005:InformationSet, DIN77005:Data, DIN77005:Document)}
    \item ODP \textit{VDI 2206} for structural decomposition %\nolinkurl{(VDI2206:MechatronicSystem, VDI2206:Module, VDI2206:Component)}
    \item  ODP \textit{VDI 3682} for functions and product, energy or information flows %(\nolinkurl{VDI3682:Process} consisting of \nolinkurl{VDI3682:ProcessOperator} with inputs/outputs \nolinkurl{VDI3682:Product}, \nolinkurl{VDI3682:Energy}, \nolinkurl{VDI3682:Information})
    \item ODP \textit{UML State Machine} for discrete behavior description %\nolinkurl{(UMLSM:State, UMLSM:Transition, UMLSM:Event, UMLSM:Action)} 
    \item ODP \textit{OpenMath} for continuous behavior description with mathematical equations %\nolinkurl{(OM:Application, OM:Operator, OM:Object)}
    %\item  ODP \textit{SOSA/SSN} for observations and actuations \nolinkurl{(SOSA:Sensor, SOSA:ObservableProperty, SOSA:Actuator, SOSA:ActuableProperty)}
    \item ODP \textit{DIN 17359} for condition monitoring and fault diagnosis %\nolinkurl{(e.g. DIN17359:Parameter, DIN17359:Anomaly, DIN17359:Fault)}
    % \item ODP \textit{DINEN61360} for describing characteristics of system information \nolinkurl{(DINEN61360:DataElement, DINEN360:TypeDescription and DINEN61360:InstanceDescription)}
\end{itemize}

The \texttt{CPSMod} alignment ontology is intentionally kept generic across CPS domains. Domain-specific schemas  such as DEXPI for process engineering can be aligned with \texttt{CPSMod} through its modular structure as a complementary extension.

The Graph RAG pipeline (depicted in dark green in Figure \ref{fig:framework}) maps data and digital models from the DPPT into the A-Box of the knowledge graph and keeps them aligned with the T-Box. Mapping languages such as RML and R2RML should be used to convert tables, engineering artifacts, simulation outputs, and condition monitoring results into RDF. The populated ontology is stored in a graph database and exposed through SPARQL 1.1. In order to interact with the established knowledge graph, a registry of predefined SPARQL templates (\nolinkurl{SELECT}/\nolinkurl{CONSTRUCT}/\nolinkurl{ASK}/\nolinkurl{DESCRIBE}) is included. Agents invoke these templates to fetch compact, relevant subgraphs that are appended as structured context to augment the prompt for the LLM. The interpretation of this knowledge graph is guided by a structured prompt that explicitly references the underlying T-Box, thereby specifying how entities and relations are to be understood and which semantics are relevant for the current task. %In combination with an explicit definition of agent roles and tasks this design ensures that the LLM does not merely access isolated facts but reasons over coherently structured context.
Graph RAG is used because recovery requires traversing relations among equipment, functions, states, faults, guards, actuators, and limits, rather than retrieving isolated document chunks. 
%DEXPI was not used and P\&ID-relevant information was represented through CPSMod-aligned structural, functional, flow, equipment, and actuator relations. 

\subsection{Agent layer} \label{Agent}
With regard to RQ2, we introduce a structured, feedback-driven method in the \textit{Agent Layer} that iteratively refines recovery plans and control actions to reduce human intervention while maintaining operational safety. To this end, operator responsibilities are decomposed into distinct, interacting agents that reflect key cognitive functions in active FTC. The method incorporates a \textit{Monitoring}, a \textit{Planning Agent}, an \textit{Action Agent}, a \textit{Simulation Agent}, a \textit{Validation Agent}, and a \textit{Reprompting Agent}. These agents combine conventional deterministic software logic (e.g., rule execution, simulation orchestration, and constraint checks) with AI-based methods (e.g. rule-based reasoning, machine learning, and LLMs) whenever non-trivial cognitive capabilities such as interpretation, inference, or decision synthesis are required. 

LLMs are used primarily in the \textit{Planning Agent} and the \textit{Action Agent} because these roles require interpreting semantically enriched information and synthesizing of constraint-compliant recovery decisions rather than performing mere classification. The LLM is not intended to replace graph search, regulatory control, simulation, or safety validation. Rather, it acts as a supervisory candidate generator when recovery requires interpreting 
retrieved plant context and proposing discrete or continuous corrective actions. This regime is relevant when (i) symptom and fault descriptions must be interpreted over natural-language ontology annotations rather than over enumerated rules, (ii) recovery requires continuous setpoint trade-offs under interacting constraints rather than the selection of a discrete shortest path, (iii) operator-readable rationales are required for human oversight, or (iv) the framework must be extensible to new faults, sensors, or actuators through knowledge-graph augmentation rather than through controller re-design. %\rev{The LLM is not intended to replace graph search, regulatory control, simulation, or safety validation; rather, it acts as a supervisory candidate generator when recovery requires interpreting retrieved plant context and proposing discrete or continuous corrective actions.} %Concretely, the Planning Agent must infer a feasible recovery path under the active fault and current process state, while the Action Agent must translate this high-level plan into actuator-level commands that respect admissible operating envelopes and procedural logic. 
In addition, LLMs can articulate intermediate rationales for each decision, which supports post-hoc interpretation, validation, and human oversight.

\paragraph{\textbf{{Monitoring Agent:}}} 
The \textit{Monitoring Agent} acquires the process context of each control loop and sequence. In the present implementation, it is not LLM-based, but a deterministic/data-driven component for processing synchronized sensor, actuator, alarm, and event signals. It ingests process values (PVs), setpoints (SPs), manipulated variables (MVs), actuator commands, and actuator feedback, covering both discrete states and continuous signals.

The agent detects faults as deviations from nominal behavior using statistical baselines (e.g., change-point tests, residual monitoring) or data-driven models (e.g., autoencoders, isolation forests). If no fault is detected, normal operation continues; otherwise, the downstream recovery workflow is triggered and the plant state is set to \nolinkurl{"FAULT"}. After a validated corrective action is applied, expected transient deviations are not treated as new faults unless hard safety limits are violated.

\paragraph{\textbf{Planning Agent:}}
 The \textit{Planning Agent} receives the typed fault symptom and fault produced by the \textit{Monitoring Agent} and transforms them into a recovery plan with feasible corrective control actions. The agent is grounded in the plant’s control context by drawing on information from the ontology described in Sec. \ref{GraphRAG}. It retrieves a knowledge graph extract that captures the relevant system functions (in open-loop and closed-loop scenarios) and its supervisory behavior under normal and faulty conditions, represented by a state machine. This subgraph serves as the authoritative context for subsequent reasoning and constrains the decision space to the admissible corrective control actions supported by the plant.

The retrieval pipeline proceeds as follows. The \textit{Planning Agent} issues \nolinkurl{SPARQL} queries to obtain (i) the implicated \nolinkurl{VDI3682:Process} instance(s) and (ii) the associated \nolinkurl{UMLStateMachine:StateMachine} fragment comprising \nolinkurl{UMLStateMachine:State} and \nolinkurl{UMLStateMachine:Transition} entities, including their guards, actions, and references to operating limits. Results are returned via a \texttt{SPARQL CONSTRUCT}, yielding a self-contained subgraph suitable for planning. Conditioned on this input, the prompt instructs the LLM to traverse the state-machine graph, enumerate guard-satisfying paths from the current (faulty) state to one or more target operational states, and propose a minimal-risk recovery path subject to operating constraints.

\paragraph{\textbf{Action Agent:}}
The \textit{Action Agent} consumes the feasible state-machine paths produced by the \textit{Planning Agent} and synthesizes corrective control actions that can move the plant from the current faulty state toward an admissible operational state. Like the \textit{Planning Agent}, it is grounded in the plant’s control context via Graph RAG. In addition to the proposed paths, it queries the knowledge graph to obtain a structure–function–actuation subgraph that links processes and behavioral paths to concrete control hardware. The agent issues symptom- and path-conditioned \nolinkurl{SPARQL} queries to retrieve the implicated \nolinkurl{VDI3682:Process} and its \nolinkurl{VDI3682:ProcessOperator} elements, and to align them with the corresponding \nolinkurl{VDI2206:MechatronicSystem}, \nolinkurl{VDI2206:Module}, and \nolinkurl{VDI2206:Component} instances that realize the functionality. This alignment exposes the relevant actuation interfaces (e.g., \nolinkurl{VDI3682:Actuator} linked to modules/components), admissible commands, and operating limits. The retrieved subgraph is used as context for prompting and constrains the action search to what the plant actually allows.

The prompt instructs the LLM to propose action sequences, including discrete commands and continuous setpoint adjustments, that enable the next transition(s) on the chosen path by satisfying guards and permissives, respect limits, rate constraints, and interlocks. As a result, the LLM must provide command semantics (e.g., open V201 to 30\%, start P301 at 50\% n). These information are returned as an ordered action set per feasible path. 

\paragraph{\textbf{Simulation Agent:}}
The \textit{Simulation Agent} consumes the set of corrective control actions produced by the \textit{Action Agent} together with the current fault mode and invokes the simulation service of the DPPT. Its purpose is to test whether the proposed corrective control actions can drive the system from the current faulty state to a valid operational state under the plant’s dynamics.

The agent prepares a simulation job that includes the fault-mode specification, initial conditions and control context (PVs, SPs, MVs, modes), the action sequence with timing and ramp profiles, relevant interlocks and envelopes to be enforced. The simulation service then executes the job and returns the resulting trajectories and end states, i.e., updated PV/SP/MV values and the resulting discrete plant mode/state after applying the proposed actions under the injected fault. If the run terminates in an operationally acceptable state without critical constraint violations, the plan is flagged as dynamically feasible. Otherwise, the service returns a structured failure report (e.g., violated constraints, unsafe modes reached, unmet terminal conditions) that is propagated back to the downstream agents to trigger refinement or reprompting.

\paragraph{\textbf{{Validation Agent:}}}

The \textit{Validation Agent} decides whether the proposed action set is safe and acceptable before any actuation on the plant is executed. It takes as input the candidate actions from the \textit{Action Agent} and the structured simulation outputs (trajectories, end states, and constraint-violation flags). Because these outputs are machine-readable, the agent relies primarily on deterministic computational logic (i.e., rule- and metric-based checks) to evaluate feasibility and safety. Its first priority is to verify that the target operational state is reached without violating interlocks, permissives, or operating envelopes. Concretely, it evaluates criteria such as (i) continued satisfaction of interlocks and permissives with no limit violations, (ii) recovery time below a specified threshold, (iii) loop-performance bounds on overshoot, settling time, and steady-state error, and (iv) compliance of actuator commands with rate, travel, and duty-cycle limits.

The agent outputs a concise decision: \textit{Approve} (safe and acceptable) with the selected action set, or \textit{Revise} (not acceptable) with structured feedback identifying the blocking checks (e.g., envelope breach, excessive recovery time, unmet guard, actuator-rate violation) so downstream agents can refine the proposal. If no safe plan emerges within the latency budget, control is handed over to the safety system.

\paragraph{\textbf{Reprompting Agent}}
The \textit{Reprompting Agent} coordinates the iterative refinement loop when a recovery path or an action set is deemed unsafe or unacceptable. It consumes (i) structured blocking feedback from the \textit{Validation Agent} (e.g., envelope breaches, unmet guards, excessive recovery time, actuator-rate violations) and (ii) simulation outputs (e.g., constraint violations, signature mismatches, sensitivity indicators). When the expected failure modes of the \textit{Planning} and \textit{Action Agents} are known, the reprompting mechanism can rely on predefined feedback templates. In this case, the \textit{Reprompting Agent} selects an appropriate text block based on the structured violations returned by the \textit{Validation Agent} and inserts it into the next planning or action prompt so the upstream agent can correct the specific issue that previously caused rejection. If such correction patterns are not available a priori, because plausible causes are unknown, too diverse, or highly context-dependent, the \textit{Reprompting Agent} can instead use an LLM to generate the refinement text directly from the observed violations. The generated feedback is then included in the subsequent prompt to make explicit what failed, why it failed, and which aspect must be revised in the next iteration. Accordingly, the agent routes that feedback to the responsible agent: planning refinement feedback to the \textit{Planning Agent} (e.g., revise the recovery path, explore alternative admissible \nolinkurl{UMLStateMachine:Transition} sequences) or action refinement feedback to the \textit{Action Agent} (e.g., re-parameterize setpoint profiles, add ramp/hold/purge steps, tighten tolerances to satisfy interlocks and envelopes).

%% file: 05_CaseStudies.tex
\section{Case studies}

\subsection{Mixing module}
\begin{figure}[pos=ht]
    \centering
    \includegraphics[width=0.5\textwidth]{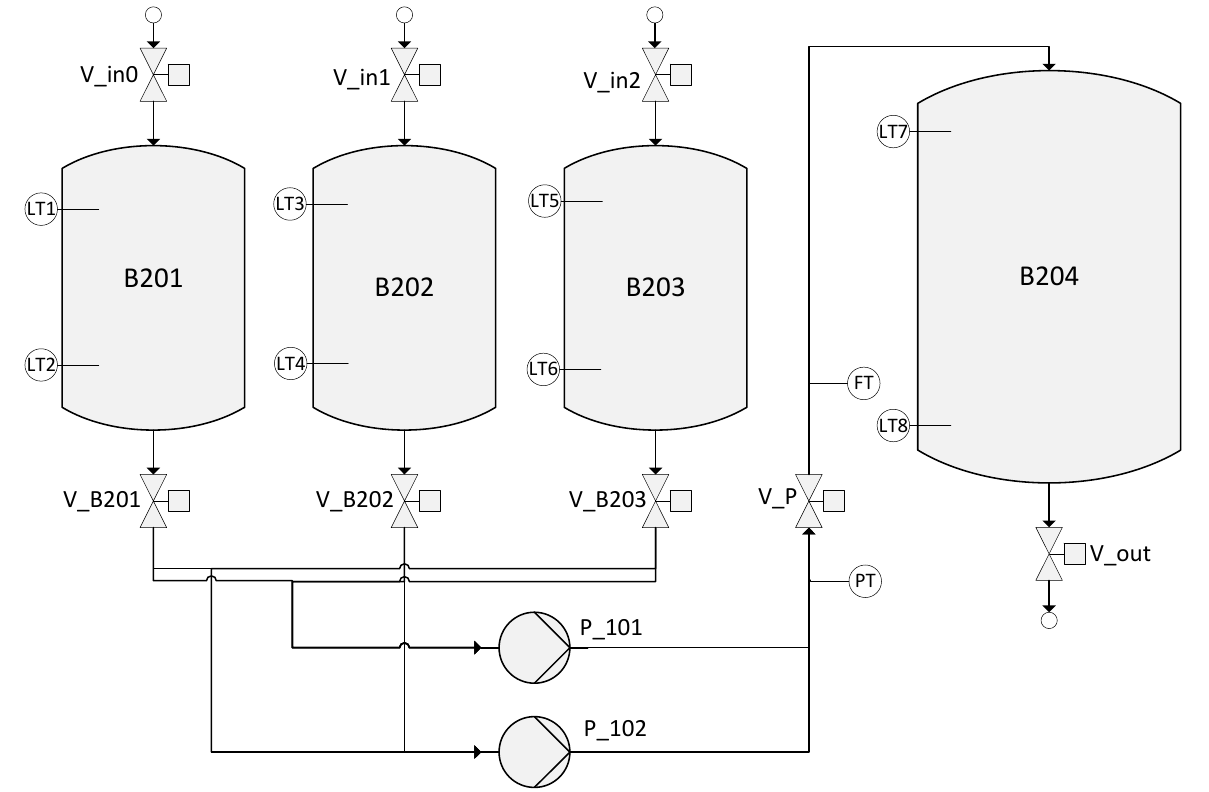}
    \caption{\textbf{Mixing Module Schematics.}}
    \label{fig:mixing_module}
\end{figure}

The first case study uses the Festo Mixing Module (see Figure \ref{fig:mixing_module} ) from the Process Automation Laboratory at Helmut Schmidt University Hamburg as a reference and realizes the laboratory setup in a Python-based simulation. The simulation mirrors the laboratory setup and provides all information required to instantiate and evaluate the proposed framework. 

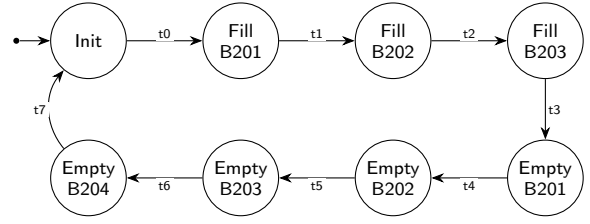
\begin{figure}[pos=t]
  \centering
  \begin{tikzpicture}[->,>=Stealth,node distance=8mm and 10mm,
      every node/.style={font=\scriptsize},
      state/.style={circle,draw,minimum size=10mm,inner sep=0pt,align=center},
      trans/.style={font=\tiny,fill=white,inner sep=1pt}]
    
    \node[state] (Init) {Init};
    \node[state,right=of Init] (F1) {Fill\\B201};
    \node[state,right=of F1] (F2) {Fill\\B202};
    \node[state,right=of F2] (F3) {Fill\\B203};
    
    \node[state,below=of F3] (E1) {Empty\\B201};
    \node[state,left=of E1] (E2) {Empty\\B202};
    \node[state,left=of E2] (E3) {Empty\\B203};
    \node[state,left=of E3] (E4) {Empty\\B204};
    
    \node[fill=black,circle,inner sep=0.8pt,left=4mm of Init] (start) {};
    \draw[->] (start) -- (Init);
    
    \path
      (Init) edge node[trans,above]{t0} (F1)
      (F1)   edge node[trans,above]{t1} (F2)
      (F2)   edge node[trans,above]{t2} (F3)
      (F3)   edge node[trans,right]{t3} (E1)
      (E1)   edge node[trans,below]{t4} (E2)
      (E2)   edge node[trans,below]{t5} (E3)
      (E3)   edge node[trans,below]{t6} (E4)
      (E4)   edge[bend left=40] node[trans,left]{t7} (Init);
  \end{tikzpicture}
  \caption{\textbf{State machine of the Mixing Module under normal operation, 
with eight states (Init, Fill B201--B203, Empty B201--B204) and 
transitions t0--t7.}}
  \label{fig:mixer-fsm-cycle}
\end{figure}

It comprises four connected tanks (\nolinkurl{B201}–\nolinkurl{B204}), a central pump (\nolinkurl{P101}), a bypass pump (\nolinkurl{P102}) and controllable on–off valves (for example \nolinkurl{V\_in0}) that serve as actionable control inputs for the agents. During operation, liquid is filled into \nolinkurl{B201}–\nolinkurl{B203} and subsequently transferred to \nolinkurl{B204} in a predefined sequence. Mode transitions are governed by discrete control logic that reacts to process conditions (see Figure \ref{fig:mixer-fsm-cycle}). Boolean level thresholds (e.g., \nolinkurl{sensor_discrete_tank_B203_high (LT5)}) trigger mode changes once predefined fill heights are reached. In addition, continuous measurements from pressure sensors (e.g., \nolinkurl{sensor_continuous_pressure (PT)}), a volumetric flow sensor (\nolinkurl{sensor_continuous_volumeFlowRate (VT)}), and the pump-speed reference enable precise control and support anomaly detection by comparison to expected operating profiles. The simulation supports fault injection for \texttt{clogging}, \texttt{leakage}, \texttt{pump\_degradation}, \texttt{pump\_failure}, and \texttt{sensor\_fault}. 

The agent framework operates by interpreting the current process state, including tank levels and fault conditions, to determine appropriate control actions. For planning, the state machine representing the plant's behavior is stored in the knowledge graph. A \texttt{SPARQL CONSTRUCT} query extracts the relevant subgraph (states, transitions, and transition guards), capturing operational constraints such as fault-dependent path selection (e.g., via bypass). Based on this semantic context and the current fault condition, the \textit{Planning Agent} selects the next target state. The target state is then passed to the \textit{Action Agent}, which queries the knowledge graph for the actuator configuration associated with that state and returns the set of actuators to be activated. 

\subsection{Continuous stirred-tank reactor}
\begin{figure}[pos=t]
    \centering
    \includegraphics[width=0.5\textwidth]{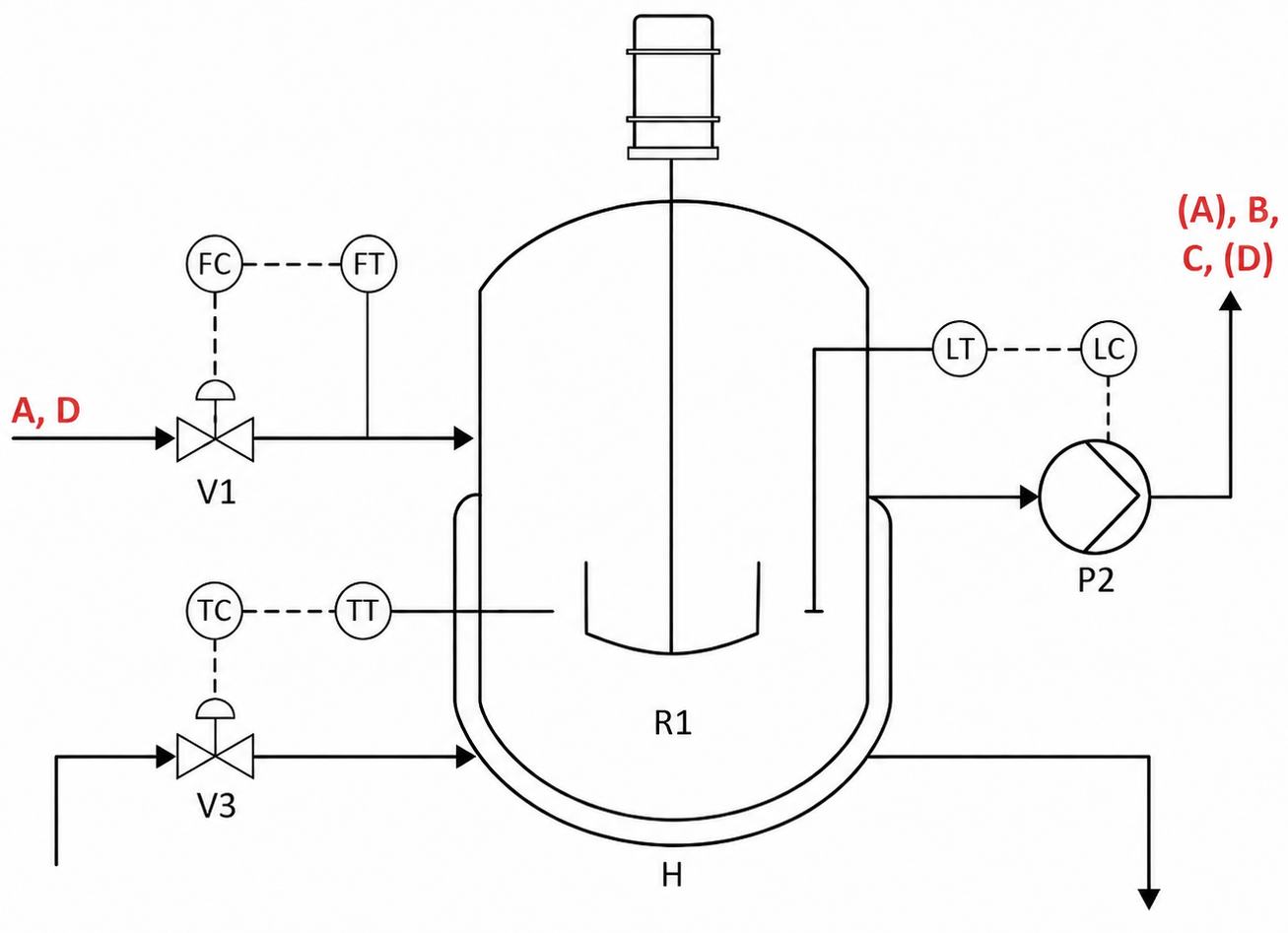}
    \caption{\textbf{CSTR Schematics.}}
    \label{fig:cstr_flowdiagram}
\end{figure}

For the second case study, we draw on the continuous stirred-tank reactor (CSTR) (see Figure \ref{fig:cstr_flowdiagram}), implemented as a Python-based simulation. The CSTR is widely used to evaluate modeling and control methods and is well established in process systems engineering and process control communities (\cite{Markaj.2024}). The system consists of a single reactor vessel \nolinkurl{R1} with variable liquid level, an inlet valve \nolinkurl{V1}, an outlet pump \nolinkurl{P2}, and a cooling jacket \nolinkurl{H1} supplied via a cooling valve \nolinkurl{V3}. Two reactants, A and D, are fed into the reactor, where A reacts with D to form B while simultaneously undergoing a parallel conversion to C. The reactor temperature \nolinkurl{T1} and the liquid level \nolinkurl{L} are measured continuously and constitute the primary controlled variables. Using the simulation, the following fault scenarios can be injected and analyzed: \texttt{fouling}, \texttt{pump\_degrade}, and \texttt{cool\_stuck\_closed}.

The CSTR is operated under regulatory control using three interacting PID control loops. The inlet-flow control loop regulates the feed flow by manipulating the inlet valve \nolinkurl{V1}. The level control loop regulates the reactor liquid level by manipulating the outlet flow via pump \nolinkurl{P2}, thereby compensating for deviations between inlet and outlet flow. The temperature control loop maintains the reactor temperature at a specified setpoint by adjusting the coolant flow through valve \nolinkurl{V3}, which determines the heat removal via the jacket \nolinkurl{H1}. All controllers are implemented as standard PID blocks and interact through the coupled mass and energy balances of the reactor.
To enable condition monitoring, statistical reference parameters (mean, standard deviation, percentile-based thresholds, and three-sigma limits) were derived a priori from fault-free operation for key monitoring variables, including measured temperature and level, inlet and outlet flows, cooling flow, actuator positions, balance residuals, and temporal gradients. Deviations from these reference ranges are interpreted as fault symptoms and linked to fault descriptions used by the \textit{Planning Agent}. Based on the identified symptom and the finite-state machine representation of the process behavior, the \textit{Planning Agent} selects the target operating state. The \textit{Action Agent} then determines suitable setpoint adaptations for the control loops to safely continue operation under the detected fault condition. The regulatory control loops, signal, material, and energy flows, as well as their transformations, are explicitly depicted in the knowledge graph. In addition, the underlying mathematical relations for every CSTR function are represented to provide the LLM agents with causal awareness of the control structure and process dynamics. The knowledge graph for both studies is stored in a graph data base (Ontotext GraphDB).

\subsection{Evaluation criteria} \label{sec:metrics}

With regard to the RQs introduced at the beginning, three overarching properties of the proposed LLM-agent framework are evaluated: \textit{decision accuracy}, \textit{added latency}, and \textit{computational cost}.

\textit{Accuracy} is assessed using four metrics capturing whether the agents synthesize safe and effective recovery decisions: (i) \textit{end-to-end success rate}, (ii) \textit{recovery rate}, (iii) \textit{plan correctness}, and (iv) \textit{action correctness}. The \textit{success rate} indicates whether a run reaches the stabilized terminal condition without safety violations and without exceeding the reprompt limit. The \textit{recovery rate} captures the fraction of reprompted runs that can be successfully corrected. \textit{Plan correctness} assesses whether the \textit{Planning Agent} selects recovery paths and state transitions consistent with the formal process logic (admissible transitions/guards) and the observed plant state. \textit{Action correctness} assesses whether the \textit{Action Agent} implements the intended recovery decision while respecting admissible control constraints. Its operationalization depends on the case-study control character (binary actuator commands vs.\ continuous setpoint adaptation) and is specified in the corresponding results subsections.

\textit{Latency} addresses how rapidly safe corrective control actions can be produced. It is measured as both \textit{per-iteration decision time} and \textit{total wall-clock time} from fault onset (or detection) to successful stabilization (evaluated corrective control action), thereby separating local decision responsiveness from global recovery time.

\textit{Costs} are measured primarily via token consumption ( \textit{input token} and \textit{output token}) aggregated over a complete run and disaggregated by agent roles, with additional overhead attributed to reprompting triggered by invalid or infeasible intermediate outputs. 
% \end{enumerate}

For an exploratory study, each  fault scenario was executed $n=10$ times per model configuration to assess the effectiveness and efficiency of the proposed LLM-agent framework for active FTC. Prior to fault evaluation, we ran $n=10$ fault-free baseline trials per model for both the CSTR and the Mixing Module, confirming that the agent system can reliably steer nominal operation in both cases. The subsequent evaluation therefore focuses exclusively on fault-handling behavior. 

\subsection{LLM selection}
The primary contribution of this work lies in the proposed framework architecture rather than a comprehensive LLM benchmark. Consequently, model selection was guided by proof-of-concept requirements, namely to demonstrate that lightweight LLMs can reliably execute knowledge-grounded decision tasks within active FTC constraints.

The decision tasks addressed in this work are highly structured as relevant system knowledge, admissible actions, and operational constraints are explicitly provided via Graph RAG and task-specific prompts. Under these conditions, the ability to rapidly interpret structured context and produce consistent, constraint-compliant outputs becomes more critical than strong general reasoning capabilities. Importantly, fault localization is out of scope and assumed to be provided by the monitoring layer, as diagnosis typically requires broader evidence integration and stronger open-ended reasoning.

Based on these considerations, two models from OpenAI’s efficiency-optimized tier were selected: GPT-4o-mini and GPT-4.1-mini. This pairing enables a controlled comparison within a single vendor ecosystem, isolating the effect of architectural model improvements while holding API behavior, tokenization, and tool-calling interfaces constant. Both models represent a cost–latency operating point that is highly relevant for industrial deployment, where inference costs and response times directly impact the feasibility of LLM-assisted control architectures.

%% file: 06_Results.tex
\section{Results}

\subsection{Mixing module results}
Table~\ref{tab:mixing_results} summarizes the results for the Mixing Module. We evaluate the agent performance using the \textit{accuracy} metrics introduced in Sec.~\ref{sec:metrics}. For the Mixing Module, \textit{action correctness} is additionally quantified using \textit{action precision} and \textit{action recall} over the selected actuator set relative to a ground-truth action set (false positives vs.\ false negatives), reflecting the binary (on/off) nature of actuator commands. 

Looking at the results, GPT-4.1-mini achieved perfect performance with a 100\% \textit{success rate} across all 60 runs, requiring no corrective reprompts. All iterations exhibited flawless action execution, with both action correctness (iterations achieving 100\% correct actuator configuration) and \textit{action precision} (ratio of correct actions to total actions) at 100\%. In contrast, GPT-4o-mini succeeded in 51 of 60 runs (85\%), with failures concentrated in two scenarios: \texttt{pump\_degradation} (50\% success) and \texttt{sensor\_fault} (60\% success). The \textit{recovery rate} metric reveals that these failures stem from systematic errors rather than correctable mistakes: only 44\% of runs requiring correction in \texttt{pump\_degradation} and 33\% in \texttt{sensor\_fault} could be recovered through reprompting, compared to 100\% recovery in all other scenarios.

\begin{figure}[pos=H]
    \centering
    \includegraphics[width=1\linewidth]{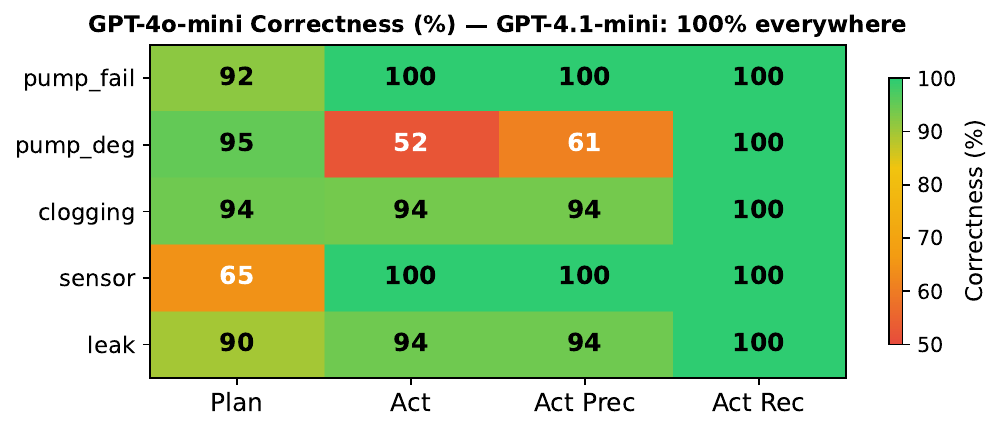}
    \caption{\textbf{Plan and action correctness per fault scenario for the Mixing Module (GPT-4o-mini), covering recovery-path correctness (Plan), actuator-configuration correctness (Act), action precision, and action recall. GPT-4.1-mini achieved 100\% across all cells.}}
    \label{Heatmap}
\end{figure}

Failure analysis identified two distinct error patterns corresponding to different agent responsibilities (see Figure \ref{Heatmap}). In \texttt{pump\_\allowbreak degradation}, the \textit{Action Agent} exhibited unnecessary activations. While correctly identifying all required actuators (100\% \textit{action recall}), it additionally activated the main pump despite the degradation condition, resulting in an \textit{action precision} of only 61.4\% (71 false positive activations out of 184 total). This represents a separation-of-concerns violation where the \textit{Action Agent }re-interpreted fault conditions that had already been resolved by the \textit{Planning Agent's} state selection. Consequently, only 52\% of iterations achieved a fully correct actuator configuration. In \texttt{sensor\_fault}, the failure originated at the planning stage, where \textit{plan correctness} dropped to 64.6\% as the model incorrectly selected bypass paths when the normal operational path remained viable. Notably, the \textit{Action Agent} performed flawlessly in this scenario (100\% action correctness, 100\% \textit{action precision}), indicating correct execution of flawed planning decisions.

A notable finding across all experiments is that \textit{action recall} was 100\% for both models across all fault scenarios: no required actuator was ever omitted. All 780 expected activations for GPT-4.1-mini and all 895 (including necessary iterations with reprompts) for GPT-4o-mini were correctly issued. This indicates that the knowledge graph grounding successfully conveyed which actuators are associated with each state. The failures made by GPT-4o-mini arose exclusively from additional, non-permitted actuator activations beyond those specified in the knowledge graph, rather than from missing required activations. This asymmetric error pattern suggests that while the semantic grounding reliably leads to necessary actions, weaker models may inject additional "common-sense" reasoning that conflicts with the formal process specification. This motivates stronger plan-conditioned action schemas that restrict the \textit{Action Agent} to the actuator set admissible for the selected recovery path.

Both models demonstrated comparable decision \textit{latencies} (GPT-4.1-mini: 2.4\,s planning, 2.2\,s action; GPT-4o-mini: 2.4\,s planning, 3.0\,s action). Mean \textit{wall-clock time} was 32.7\,s for GPT-4.1-mini and 40.7\,s for GPT-4o-mini, with the difference attributable to reprompting overhead. The remaining wall-clock time is due to graph retrieval, repeated supervisory steps, simulation rollout, validation, orchestration, logging, and reprompting overhead. \textit{Token consumption} averaged 55.8k for GPT-4.1-mini and 59.4k for GPT-4o-mini. Failed runs consumed approximately 43.1k tokens due to additional reprompt cycles before termination.

\begin{table}[pos=t]
\centering
\caption{Mixing Module operational results ($n=10$ per scenario).}
\label{tab:mixing_results}
\resizebox{\columnwidth}{!}{%
\begin{tabular}{@{}llccccc@{}}
\toprule
\textbf{Fault} & \textbf{Model} & \textbf{Succ.} & \textbf{Rec.} & \textbf{Time} & \textbf{Tok.~In} & \textbf{Tok.~Out} \\
\midrule
\multirow{2}{*}{pump\_fail.} 
 & GPT-4.1-mini& 10/10 & -- & 35.5\,s & 54.1k & 1.7k \\
 & GPT-4o-mini  & 10/10 & 100\% & 38.6\,s & 60.4k & 1.4k \\
\midrule
\multirow{2}{*}{pump\_deg.} 
 & GPT-4.1-mini & 10/10 & -- & 34.0\,s & 54.2k & 1.7k \\
 & GPT-4o-mini   & \textbf{5/10} & 44\% & 33.2\,s & 43.2k & 1.2k \\
\midrule
\multirow{2}{*}{clogging} 
 & GPT 4.1-mini & 10/10 & -- & 33.7\,s & 54.2k & 1.8k \\
 & GPT-4o-mini  & 10/10 & 100\% & 50.5\,s & 60.5k & 1.5k \\
\midrule
\multirow{2}{*}{sensor} 
 & GPT-4.1-mini& 10/10 & -- & 30.7\,s & 54.1k & 1.6k \\
 & GPT-4o-mini  & \textbf{6/10} & 33\% & 42.4\,s & 63.1k & 1.5k \\
\midrule
\multirow{2}{*}{leak} 
 & GPT-4.1-mini & 10/10 & -- & 32.2\,s & 54.2k & 1.8k \\
 & GPT-4o-mini  & 10/10 & 100\% & 45.0\,s & 63.7k & 1.6k \\

\midrule\midrule
\multirow{2}{*}{\textbf{All}} 
 & GPT-4.1-mini & \textbf{60/60} & \textbf{--} & 32.7\,s & 54.1k & 1.7k \\
 & GPT-4o-mini  & 51/60 & 80\% & 40.7\,s & 58.0k & 1.4k \\
\bottomrule
\end{tabular}%
}
\vspace{1mm}
\par\raggedright\scriptsize
\textit{Succ.}=runs completing without errors; 
\textit{Rec.}=recovery rate (successful corrections / runs needing correction, -- = none needed); 
\textit{Time}=mean wall-clock time; 
\textit{Tok.~In/Out}=input/output tokens ($\times 10^3$).
\end{table}

\subsection{CSTR results}
Table~\ref{tab:cstr_results} summarizes the performance of the CSTR case study for three injected fault scenarios (\texttt{fouling}, \texttt{pump\_degrade}, and \texttt{cool\_stuck\_closed}).
In contrast to the Mixing Module, the CSTR recovery task is continuous-control dominated. Once the \textit{Monitoring Agent} triggers an intervention during normal operation  because of a fault being detected, the \textit{Action Agent} must propose corrected setpoints $(T_{sp},L_{sp},F_{in,sp})$ that satisfy safety constraints and recover regulation despite actuator limitations.

A run is counted as successful only if a validated LLM proposal is applied without exceeding the reprompt limit. Concretely, \textit{success rate} requires that (i) a setpoint intervention was applied, (ii) the validation passed, and (iii) the run did not hit the configured reprompt limit (here $5$). Importantly, runs that rely on the deterministic fallback policy after exceeding the reprompt limit are counted as failures (even though the plant may remain operational). This choice aligns the metric with the paper’s objective, demonstrating that safe recovery is achieved by the LLM-generated control action rather than by a hard-coded controller.

GPT-4.1-mini achieved a \textit{success rate} of $28/30$ (93.3\%) across all runs, while GPT-4o-mini achieved $23/30$ (76.7\%).
Both models solved \texttt{pump\_degrade} reliably ($10/10$), indicating that moderate loss of pumping effectiveness can be compensated with stable setpoint adaptations under the given acceptance criteria.

The dominant failure mode arises under \texttt{cool\_stuck\_\allowbreak closed}, which emulates a cooling limitation (cooling valve constrained near the closed region).
Here, GPT-4.1-mini succeeded in all runs ($10/10$), typically requiring reprompting (mean $\approx 2$ reprompts) to converge to a safe setpoint triple.
In contrast, GPT-4o-mini succeeded in only $4/10$ runs, with failures largely attributable to hitting the reprompt budget, i.e., repeated proposals that did not satisfy the rollout recovery requirement within the acceptance thresholds.

Across successful and failed runs, GPT-4.1-mini shows lower mean per-call \textit{action latency} (2.1\,s vs.\ 2.7\,s).
Token consumption per run is comparable across models, with mean \textit{Action Agent} input on the order of $\sim$13–15k tokens and output on the order of a few hundred tokens, reflecting that the prompt is dominated by the structured snapshot and knowledge graph context while the completion is constrained to a short JSON response. Compared to the Mixing Module, overall token usage per run is lower because the agent workflow is invoked less frequently. In the Mixing Module, the agents are called at each discrete supervisory step whenever a state transition is required, whereas in the CSTR case the decision loop is typically triggered only once per run.

\begin{table}[pos=htbp]
\centering
\caption{CSTR operational results ($n=10$ per scenario).}
\label{tab:cstr_results}
\resizebox{\columnwidth}{!}{%
\begin{tabular}{@{}llcccccc@{}}
\toprule
\textbf{Fault} & \textbf{Model} & \textbf{Succ.} & \textbf{Rec.} & \textbf{Time} & \textbf{Act.Lat} & \textbf{Tok.~In} & \textbf{Tok.~Out} \\
\midrule
fouling & GPT-4.1-mini & 8/10 & 43\% & 50.5\,s & 1.9\,s & 15.0k & 0.2k \\
 & GPT-4o-mini & 9/10 & 50\% & 42.7\,s & 2.3\,s & 13.7k & 0.2k \\
\midrule
pump\_degrade & GPT-4.1-mini & 10/10 & -- & 51.1\,s & 1.9\,s & 13.7k & 0.2k \\
 & GPT-4o-mini & 10/10 & -- & 36.8\,s & 2.4\,s & 11.6k & 0.2k \\
\midrule
cool\_stuck\_closed & GPT-4.1-mini & 10/10 & 100\% & 54.0\,s & 2.4\,s & 13.5k & 0.3k \\
 & GPT-4o-mini & \textbf{4/10} & \textbf{0}\% & 50.5\,s & 3.5\,s & 14.9k & 0.3k \\
\midrule\midrule
\textbf{All} & GPT-4.1-mini & \textbf{28/30} & 75\% & 51.9\,s & 2.1\,s & 14.1k & 0.2k \\
 & GPT-4o-mini & 23/30 & 50\% & 43.3\,s & 2.7\,s & 13.4k & 0.2k \\
\bottomrule
\end{tabular}%
}
\vspace{1mm}
\par\raggedright\scriptsize
\textit{Succ.}=runs in which a validated LLM action was applied without exceeding the reprompt limit (fallback runs count as failures);
\textit{Rec.}=recovery rate among runs that required at least one reprompt (successful / reprompted);
\textit{Wall}=mean wall-clock time per run; \textit{Act.Lat}=mean per-call Action Agent latency;
\textit{Tok.~In/Out}=mean input/output tokens for the action agent ($\times 10^3$).
\end{table}

\begin{figure}[pos=t]
\centering
\begin{subfigure}{\columnwidth}
  \centering
  \includegraphics[width=\columnwidth]{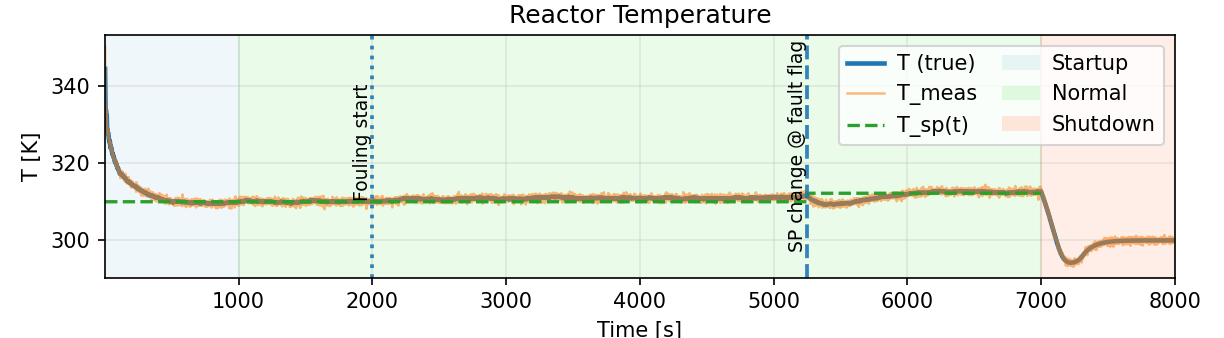}
  \caption{Reactor temperature $T$ and setpoint $T_{sp}(t)$. The dashed marker indicates the validated LLM intervention; setpoints are masked during shutdown.}
  \label{fig:cstr_temp}
\end{subfigure}

\vspace{1.2mm}

\begin{subfigure}{\columnwidth}
  \centering
  \includegraphics[width=\columnwidth]{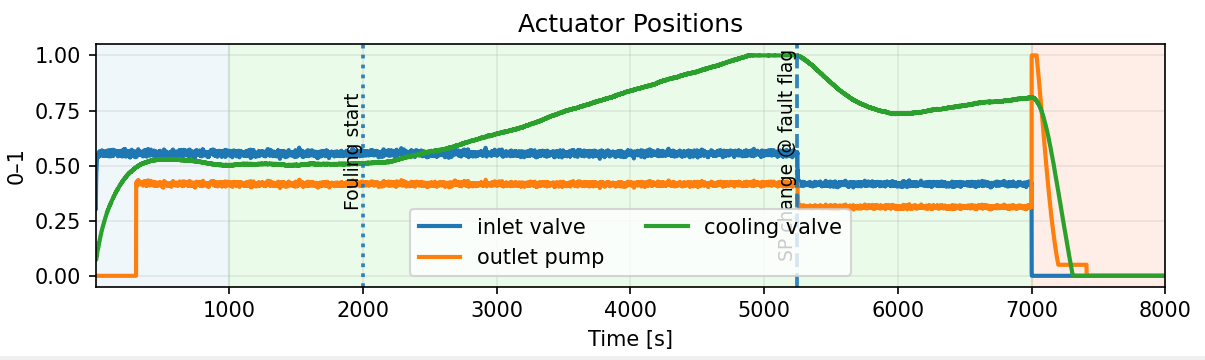}
  \caption{Actuator positions during recovery, illustrating constrained cooling and coordinated valve/pump adjustments.}
  \label{fig:cstr_act}
\end{subfigure}

\caption{\textbf{Action Agent setpoint intervention for temperature recovery in the CSTR.}} %After fault detection, the Action Agent proposes a validated setpoint intervention that restores temperature regulation through coordinated actuator adjustments.
% }
\label{fig:cstr_core}
\end{figure}

Under the \texttt{fouling} fault, the effective heat-transfer capability of the reactor degrades over time, reducing the ability of the cooling circuit to remove heat. 
As shown in Figure~\ref{fig:cstr_core}, this loss of heat exchange causes the cooling actuator to progressively increase toward its upper limit, while the reactor temperature continues to rise and eventually exceeds the nominal safety threshold despite maximal cooling effort. 
Once further actuation is no longer effective, recovery is achieved by reducing the inlet flow setpoint, thereby lowering the heat generation rate and restoring thermal feasibility at the cost of reduced production. 
The resulting operating point reflects a deliberate trade-off enforced by the validated LLM intervention: stable operation under degraded heat transfer rather than full nominal throughput.

%% file: 07_Discussion.tex
\section{Discussion}
Drawing on the case study results, this section addresses the RQs, interprets the findings, discusses implications for LLM-based active FTC, and highlights limitations of the proposed framework.

\textit{RQ1 (Knowledge Representation and Retrieval for LLM Grounding):} The results demonstrate that Graph RAG-based retrieval provides sufficient grounding for LLM agents to derive correct FTC decisions. In particular, GPT-4.1-mini achieved 100 \% \textit{plan} and \textit{action correctness} across all fault scenarios (apart from \texttt{fouling}), indicating that the semantic representation captured the plant-specific knowledge required for decision-making. The ontology-based representation enabled an unambiguous interpretation of process states, fault conditions, and admissible control actions by providing a shared, formally defined semantic context for both agents.

The ontology proved effective in integrating heterogeneous engineering artifacts, including P\&ID information, sensor–actuator mappings, control loops cause–effect relations, and behavioral specifications, into a unified semantic layer. By preserving relational semantics between system components, the knowledge graph allowed the \textit{Planning Agent }to identify fault-dependent constraints on operational paths and to reason about viable alternatives without requiring explicit enumeration of all fault scenarios. These findings indicate that Graph RAG is a viable mechanism for contextualizing dispersed plant knowledge and supporting first-principles reasoning in active FTC.

A key limitation lies in the effort required to construct the knowledge graph. The current implementation relies on structured inputs that, in industrial practice, are often available only in unstructured or semi-structured form. While automated extraction using LLMs or vision-language models could significantly reduce this effort, such approaches were beyond the scope of the present study.

\textit{RQ2 (Synthesis and Validation of Recovery Plans):} The \textit{Planning} and the \textit{Action Agent} successfully synthesized correct recovery plans and validated them through the simulation prior to execution. The \textit{Planning Agent} consistently selected fault-appropriate state transitions, while the \textit{Action Agent} derived actuator configurations aligned with the selected operational mode. The simulation effectively detected infeasible actions before execution, and reprompting enabled recovery from transient decision errors.

The \textit{recovery rate} metric revealed an important distinction between correctable and systematic errors. In scenarios where GPT-4o-mini failed (e.g. \texttt{sensor\_fault}: 33\% recovery for the Mixing Module, \texttt{cool\_stuck\_closed}: 0\% recovery for the CSTR), the low recovery rates indicate that reprompting within the applied limits could not resolve the underlying reasoning failure. This contrasts with other scenarios achieving 100\% recovery, where initial errors were successfully corrected through feedback. The implication is that validation and reprompting provide effective safeguards against transient errors, but they cannot compensate when the model lacks reliable fault localization. If the LLM cannot attribute the fault to a specific component or subsystem, or is not provided with localization context that constrains which recovery actions are admissible, it may repeatedly propose plausible yet inapplicable actions. A key limitation concerns the evaluation of unknown faults. The digital twin can only validate control actions against fault conditions that have been modeled a priori. Truly novel faults cannot be simulated without prior characterization. Furthermore, establishing a ground truth for planning decisions requires expert specification of expected recovery paths, which may not be available for complex or unprecedented fault combinations.

\textit{RQ3 (Effectiveness and Timeliness of Corrective Actions):} The framework demonstrated effective derivation of corrective control actions for both the Mixing Module (open-loop, sequential batch process) and the CSTR (closed-loop, continuous control). GPT-4.1-mini achieved 100\% success across all scenarios with mean wall-clock times of 32.7\,s for the mixing module, indicating that the approach generalizes across different process characteristics and control paradigms.

Per-decision latencies are acceptable for batch and supervisory control scenarios, where the process evolves slowly and interventions are needed every few seconds or minutes. However, these latencies preclude application to processes with hard real-time constraints. Safety-critical systems requiring millisecond responses would necessitate alternative architectures, such as offline policy synthesis. 

Finally, scalability remains only initially characterized. Larger plants increase the relevant decision context, including candidate actions, interacting constraints, coupled faults, and grounding requirements, so future work must evaluate how \textit{recovery success} and \textit{decision latency} scale with this complexity. The framework is feasibility-first rather than optimality-first and does not estimate globally optimal recovery paths. It also assumes a representative DPPT; deployment under model mismatch, sensor noise, actuator degradation, or unmodelled dynamics would require uncertainty-aware validation, such as robust envelopes, online calibration, or interval bounds. Cloud-hosted LLMs are used only for research evaluation; industrial deployment would require on-premise/edge models, IT/OT segmentation, access control, audit logging, prompt-injection mitigation, and deterministic fallback.

%% file: 08_SummaryAndOutlook.tex
\section{Summary and outlook}
The presented work introduced an active FTC framework combining LLM-based agents, a digital twin, and a Graph RAG knowledge layer. Specialized \textit{Planning} and \textit{Action Agents} synthesize corrective actions that are validated in simulation before execution. Structured knowledge is provided through the \texttt{CPSMod} ontology, which unifies structural, functional, behavioral, and fault-related information for agent use. Evaluation on a Modular Mixing Unit and a Continuous Stirred-Tank Reactor demonstrated safe recovery across discrete supervisory and continuous closed-loop tasks, indicating that semantically grounded LLM agents with simulation-based validation provide a promising basis for active FTC.  Future work will target automated knowledge-graph construction, larger coupled process systems, and tighter integration of fault detection/localization with recovery agents.